\documentclass{cimento}

\usepackage{graphicx}  

\begin{document}

\title{A search for directional violations of the Lorentz invariance\\through the study of a possible anisotropy of particle lifetimes}
\author{A.~de Angelis\from{ins:ud}, {M. de Maria}\from{ins:ve}, {M. Antonelli}\from{ins:fr}, {M. Dreucci}\from{ins:fr}}

\instlist{\inst{ins:ud} Max-Planck-Institut f\"ur Physik
(Werner-Heisenberg-Institut),
F\"ohringer Ring 6,
M\"unchen, Germany \thanks{On leave of absence from Universit\`a di Udine, Via delle Scienze 208, Udine, Italy.}; INFN and INAF Trieste, Italy; LIP/IST, Lisboa, Portugal
\inst{ins:ve} Universit\`a IUAV di Venezia,  Tolentini, Santa Croce 191, Venezia, Italy
\inst{ins:fr} INFN Laboratori Nazionali di Frascati, Via E. Fermi 40, Frascati, Italy
 
}

\shorttitle{Search for global anisotropies of time dilation}
\shortauthor{A. de Angelis, M. de Maria, M. Antonelli, M. Dreucci}

\PACSes{
\PACSit{03.30.+p}{Special relativity}
\PACSit{04.60.-m}{Quantum gravity}
\PACSit{95.85.Pw}{Gamma rays astronomical observations}
}

\maketitle

\begin{abstract}
From the study of a sample of about 62.3 million well reconstructed $K^0_S \rightarrow \pi^+\pi^-$ decays recorded by the KLOE detector at the DA$\Phi$NE accelerator in Frascati, the lifetimes of $K^0_S$ mesons parallel and antiparallel to the direction of motion of the Earth with respect to the Cosmic Microwave Background (CMB) reference frame have been studied. No difference has been found, and a limit on a possible  asymmetry of the lifetime with respect to the CMB has been set at 95\% C.L.:  $|A|_{\rm{CMB}} = {|\tau_{+CMB}-\tau_{-CMB}|}/{(\tau_{+CMB}+\tau_{-CMB})}< 0.98 \times 10^{-3}.$ This is presently the best experimental limit on such quantity, and it is smaller of the speed, expressed in natural units, of the Solar System with respect to the CMB $(V/c = 1.23 \times 10^{-3})$. The present limit might constrain possible Lorentz-violating anisotropical theories.
\end{abstract}

Possible violations of the Lorentz invariance have been recently suggested to explain anomalies in the propagation of cosmic rays and the transparency 
\cite{magic0} of the Universe to gamma rays \cite{cole,kifu,ste}. Violations of the Lorentz invariance might in particular imply the existence of a preferred reference  frame; this could introduce a globally anisotropical mechanics \cite{globalani}. A global anisotropy might explain a possible different behavior of photon propagation for different high energy gamma sources \cite{magic1,fermi}, and it has been recently suggested as a possible explanation of
physical observations \cite{question}. Different frameworks have been proposed \cite{21,27} to host a global anisotropy; 
widely used are the theories of Robertson \cite{12}, and Mansouri and Sexl \cite{131415}, together generally called the
RMS-theories,
and the so-called ``Standard
Model Extension'' \cite{43}.

Many experiments measuring particle lifetimes \cite{reviewrev} give evidence for a
time dilation in accordance with the Lorentz transformations,  up to the 
present experimental accuracy. Such experimental verifications do not rule out however the possibility that measured particle lifetimes depend on their direction of motion: present
isotropy tests are not very accurate \cite{deasmalska}.

Accurate isotropy tests have instead been performed about the speed of light,
and they did not give firm indications of anisotropies \cite{reviewrev,speed}. In special relativity, spatial isotropy is a crucial issue in the
synchronization procedure of two distant clocks; Einstein stressed however \cite{einstein} that
the fact that light travels at equal speeds along the opposite directions of a
particular path is ``neither a supposition nor a hypothesis about the physical
nature of light, but a stipulation'' that can be freely made so as to arrive at a
definition of simultaneity.  The so-called conventionalist thesis proposed by Reichenbach \cite{reichen} states that quantities
as the one-way speed of light are inherently conventional, and that to recognize 
this aspect is to recognize a profound feature of nature. Only proper time has
``objective status in special relativity'' \cite{friedman}; this is because one-way velocity's
value is nothing about the pattern of coincidences of
events at a given space locations, but it refers to the comparison of remote events,
and so is inevitably conventional. Thus a test of isotropy of particle lifetimes is independent of a test of the isotropy of the speed of light.

The Cosmic Microwave Background (CMB) dipole anisotropy \cite{cobe}, interpreted
as a Doppler effect, indicates the motion of the Local Group in the
direction $(\ell, b)_{\rm{CMB}}$ = $(264^\circ, 48^\circ)$ in galactic coordinates, with
a speed of $V=(369 \pm 1)$ km/s \cite{pdg}, {\em{i.e.,}} 
\begin{equation} V/c = (1.231 \pm 0.003) \times 10^{-3} \, . \label{visuc} \end{equation} 
The CMB is a unique rest frame: even if this fact does not imply by itself any anisotropy of the physical laws
(although at a small level QED should became anisotropical due to interactions with a nontrivial vacuum), 
the existence of such a natural rest frame provides a rational framework for the interpretation of any asymmetry that might possibly be discovered. The old idea of an absolute ``aether'' is exploited, with the
only difference that the preferred frame is now identified with one in which
the cosmic background radiation is locally isotropic. In fact, there is only
one frame with this property, being all other frames experiencing the dipole
anisotropy, and therefore distinguishable.

\begin{figure}
\begin{center}
\includegraphics[width=0.78\linewidth]{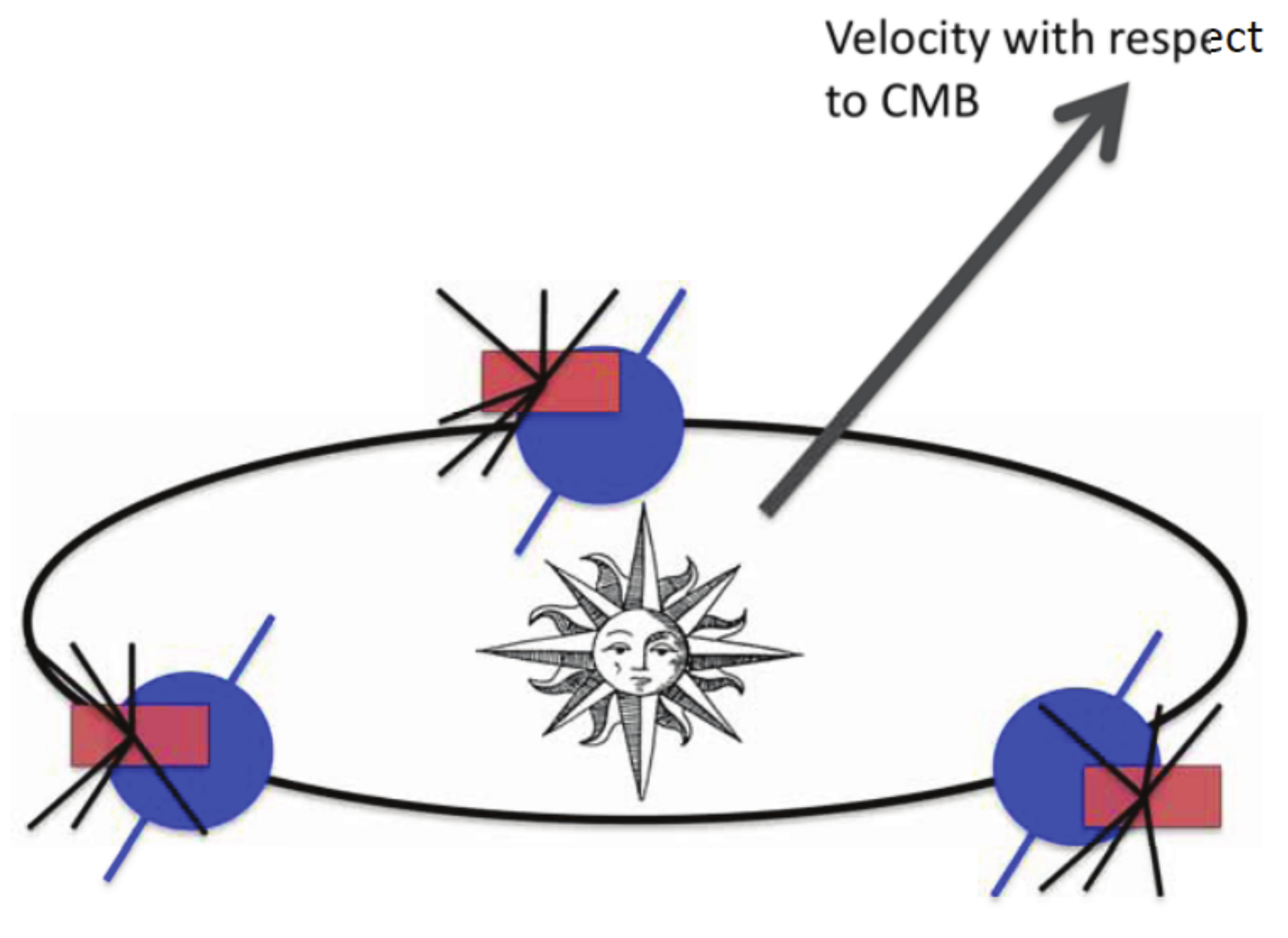} 
\end{center}
\caption{Collider detectors with $4\pi$ acceptance can be used as a probe for detecting global asymmetries: the Earth's rotation provides in the different seasons and hours of the day different orientations of the symmetry axes with respect to arbitrary directions, thus reducing detector effects.}
\label{fig:amaldilet}       
\end{figure}

Collider detectors with $4\pi$ acceptance can be used as a probe for detecting global asymmetries in the Universe \cite{deasmalska}: the Earth's rotation provides in the different seasons and hours of the day different orientations of the symmetry axes with respect to arbitrary directions, and this fact entails a strong reduction of detector effects (fig. \ref{fig:amaldilet}).

This note reports on a test of the isotropy of $K^0_S$  lifetime, which has been done \cite{michela,frascati} by comparing the lifetimes of $K^0_S$  measured by KLOE parallel and antiparallel with respect to the direction of motion of the particles with respect to the CMB system.

Selected data on $K^0_S$ decays into charged pion pairs have been used from the data collected by KLOE in 2004 and 2005. After a severe quality   selection, a total sample of about 62.3 million well reconstructed decays has been used. The $K^0_S$  
momentum\footnote{DA$\Phi$NE operates at an energy slightly above the $\phi$ mass; the $K^0_S$ have a peak momentum of about 0.11 GeV/$c$.} has been transformed from local-KLOE into galactic  coordinates \cite{michela}.  We retained only events inside a cone with an opening angle of $30^\circ$\footnote{Monte Carlo studies \cite{deasmalska} demonstrated that this
criterion maximizes the sensitivity to asymmetries. The relation between the measured
asymmetry in a cone and the asymmetry with respect to a fixed a direction $\vec{u}$ is, in case of
uniform population (which is a good approximation for the data set):
\begin{equation}\label{eq1}
A_{\vec{u}} \simeq 1.072 A_{cone} \, . \end{equation}} 
parallel $(up)$ and antiparallel $(down)$ to the direction of motion with respect to the
CMB, 
and in each cone we measured  the $K^0_S$  lifetime, $\tau,$ by a fit to the proper time distribution as in \cite{frascati}. We define the asymmetry $A_{cone}$
as 
\[ A_{cone} = \frac{\tau_{up}-\tau_{down}}{\tau_{up}+\tau_{down}} \, , \]
 finding \cite{frascati} $A_{cone,\,{\rm{CMB}}} = (-0.13 \pm 0.40) \times 10^{-3},$ consistent with zero; the error is purely statistical, since systematics largely cancel due to the definition of the asymmetry (this fact has been explicitly verified in a subsample corresponding to about 1/4 of the total statistics in \cite{frascati}). As a cross check, the asymmetry has been measured in two directions perpendicular to the direction of motion with respect to the CMB, and found again to be consistent with 
 zero\footnote{The so-called ``absolute'' direction, defined in \cite{cah04}: $(\ell, b)_{\rm{abs}}$ = $(277.6^\circ, -34.5^\circ)$ in galactic coordinates, 
 has been also studied; a value 
 $A_{cone,\,{\rm{abs}}} = (-0.1 \pm 0.3) \times 10^{-3},$ again  consistent with 0, has been found.}.

Using eq. \ref{eq1}, such a result on the asymmetry with respect to a cone translates in an
upper limit at 95\% CL referred to the direction of motion with respect to
the CMB:
\begin{equation}
 |A|_{\rm{CMB}} = \frac{|\tau_{+CMB}-\tau_{-CMB}|}{\tau_{+CMB}+\tau_{-CMB}}< 0.98 \times 10^{-3} \,  (95\% \, \rm{CL}) \, . \end{equation}

This result sets limits to non-relativistic theories, and to possible
anisotropical interactions of neutral kaons with the matter in the
universe, in an unexplored domain, improving by one order of magnitude
the previous results (\cite{deasmalska}).
The present upper limit on the asymmetry is smaller than
the relative velocity (eq. \ref{visuc}), in natural units, of the Solar System with
respect to the CMB.


\end{document}